\documentclass[a4paper, authoryear, 12pt]{elsarticle}

\usepackage[english]{babel}
\usepackage{babel}
\usepackage{fontenc}
\usepackage{graphicx}
\usepackage{amsmath,amsthm,amssymb}
\usepackage{natbib}
\usepackage[colorlinks=true,citecolor=blue]{hyperref}
\usepackage{color}
\usepackage{graphicx}
\usepackage{subfigure}
\usepackage{color}
\usepackage{newlfont}
\usepackage{multirow}
\usepackage{longtable} %Allows for long tables that stretch across many pages
\usepackage{ifthen}
\usepackage{alltt}
\usepackage{enumerate}
\usepackage{enumitem}
\usepackage[text={6.25in,8.5in},centering]{geometry}
\usepackage{tikz}
\usepackage{epstopdf}
\usepackage{float}
\usepackage{arydshln}
\usepackage[utf8]{inputenc}
\usepackage{tikz}
\usetikzlibrary{shapes,arrows}
\numberwithin{equation}{section}
\newtheorem{theorem}{Theorem}[section]

\input epsf.sty
\usepackage{lineno}
\journal{arXiv}

\begin{document}
	\title{Anticipating dengue outbreaks using a novel hybrid ARIMA-ARNN model with exogenous variables}
	
	\author[IITB]{Indrajit Ghosh}
    \author[IPM]{Shashank Gupta}
	\author[VB]{Sourav Rana\footnote{Corresponding author. Email: sourav.rana@visva-bharati.ac.in}}
	\address[IITB]{Department of Mathematics, Indian Institute of Technology, Bombay - 400076, Maharashtra, India}
    \address[IPM]{IPM (2022-27), Indian Institute of Management, Indore-453556, Madhya Pradesh, India}
    \address[VB]{Department of Statistics, Visva Bharati, Shantiniketan - 731235, West Bengal, India} 
\begin{abstract}
Dengue incidence forecasting using hybrid models has been surging in the data rich world. Hybridization of statistical time series forecasting models and machine learning models are explored for dengue forecasting with different degrees of success. In this paper, we propose a multivariate expansion of the hybrid ARIMA-ARNN model. The main motivation is to propose a novel hybridization and apply it to dengue outbreak prediction. The asymptotic stationarity of the proposed model has been established. We check the forecasting capability and robustness of the forecasts through numerical experiments. State-of-the-art forecasting models for multivariate time series data are compared with the proposed model using accuracy metrics. Dengue incidence data from San Juan and Iquitos are utilized along with rainfall as an exogenous variable. Results indicate that the proposed model improves the ARIMAX forecasts in some situations and closely follows it otherwise. The theoretical as well as experimental results reinforce that the proposed model has the potential to act as a candidate for early warning of dengue outbreaks. The proposed model can be readily generalized to incorporate more exogenous variables and also applied to other time series forecasting problems wherever exogenous variable(s) are available.
\end{abstract}


\maketitle

\section{Introduction}\label{sec1}
Dengue is the most prevalent and rapidly spreading mosquito-borne viral disease. The incidence of dengue has grown dramatically around the world in recent decades, with cases reported to WHO increasing from 505,430 cases in 2000 to 5.2 million in 2019 \cite{WHO_Dengue}. The burden of dengue has become heavier from 1990 to 2019, amidst the three decades of urbanization, warming climates and increased human mobility in much of the world \cite{Dengue_GlobalBurden}. Dengue fever is mostly observed in tropical and sub-tropical regions of the globe. In particular, several pockets in Africa, Southeast Asian countries and the western Pacific region are prone to a high burden of dengue disease. 

The virus is transmitted to humans through the bites of infected female mosquitoes, primarily the Aedes aegypti mosquito. Other species within the Aedes genus can also act as vectors, but their contribution is secondary to Aedes aegypti. While  the  majority  of  infections  are  milder asymptomatic, the more severe forms of dengue infection - dengue shock  syndrome (DSS) and dengue hemorrhagic fever (DHF) - can result in organ failure or death \cite{Dengue_DHF}. Developing a dengue vaccine has proven challenging due to various factors, such as the requirement for a tetravalent vaccine capable of providing protection against all four dengue virus (DENV) serotypes, the absence of suitable animal models for testing, and concerns surrounding the potential immune enhancement caused by the vaccine, similar to what occurs during natural infection \cite{Dengue_Vaccine}. Despite the substantial global demand, these obstacles have hindered the progress of dengue vaccine development. There is also no ready-to-use medicine for the disease. Therefore, it is of utmost importance to get some idea about future trends of dengue cases in the population.

The effectiveness of preventive measures against dengue fever is greatly enhanced by the presence of a precise early warning system that can predict upcoming epidemics. It has been established that early detection of cases and treating them can significantly reduce fatal complications \cite{degallier2010toward}. Early warning systems or forecasting models can inform the expected number of dengue cases over the coming months. This information can then be utilized to allocate resources to high-risk zones and awareness campaigns can be performed to flatten the expected dengue incidence curve \cite{world2005using,thomson2008seasonal}. Thus, public health authorities rely on model predictions for optimal management of future dengue cases. Due to the high importance of accurate forecasts of future dengue cases, many researchers have attempted this problem with different levels of success \cite{gharbi2011time,buczak2018ensemble,lauer2018prospective}. However, there is a diverse range of models that are used for dengue prediction problems, namely, compartmental SIR-type models \cite{racloz2012surveillance}, statistical time series models \cite{johnson2018phenomenological}, machine learning models \cite{guo2017developing} and ensemble models  \cite{yamana2016superensemble,deb2022ensemble}. Researchers have seen that statistical and machine learning models have achieved a higher degree of success for epidemic forecasting than deterministic compartmental models 
\cite{chakraborty2019forecasting, yamana2016superensemble, deb2022ensemble}.

The dengue forecasting problem has been tried by many statistical and machine learning models. Univariate time series of dengue cases based on historical data are modelled in numerous studies \cite{luz2008time, promprou2006forecasting}. Auto-regressive integrated moving average model (ARIMA) is mostly used to build upon past data of the series to produce future forecasts. While classical time series models are still performing well, some advanced machine learning models are also employed to forecast dengue cases \cite{guo2017developing,li2021long}. A combination of two or more forecasting models has also been developed to tackle the problem. For example, \cite{deb2022ensemble} proposed an ensemble vector auto-regressive structure combining ARIMA, negative binomial regression and generalized linear auto-regressive moving average model to predict dengue cases. The Seasonal ARIMA model is also used to study the effect of climate variables on dengue cases \cite{johansson2016evaluating}. Generalized additive models with lags in the number of cases and climatic variables are also studied for dengue forecasting \cite{baquero2018dengue}. The Auto-Regressive Neural Network (ARNN) model and its hybrid versions are also used to forecast dengue cases \cite{zhao2020machine, baquero2018dengue, chakraborty2019forecasting}. \cite{DHF_annualForecast_Thailand} developed statistical models that use biologically plausible covariates, observed by one particular month each year, to forecast the cumulative DHF incidence for the remainder of the year. They found that functions of past incidence contribute most strongly to model performance, whereas the importance of environmental covariates vary regionally. A model using time series analysis was devised to depict the pattern of dengue fever cases in Kaohsiung City \cite{Weather-predictor_Kaohsiung}. The findings revealed a statistical correlation between the incidence of dengue fever and temperature, as well as relative humidity. The most significant impact was observed at a lag of 2 months. Recently, Panja et al. \cite{panja2023ensemble} studied the effect of rainfall on dengue incidence forecasting and proposed a novel hybrid model. They found that the rainfall data indeed improves the forecasting capability of some forecasting models. 

In this paper, our main objective is to propose a multivariate extension of the hybrid ARIMA-ARNN model \cite{chakraborty2019forecasting} for dengue case forecasting using rainfall as an exogenous variable. A realistic splitting technique of the dataset will be used to check the model performance. State-of-the-art forecasting models will be compared with the proposed model using appropriate accuracy measures.

The rest of the paper is organized as follows: we present the proposed model and discussed their asymptotic stationarity in Section \ref{sec2} and Section \ref{Sec:Asmp_Stationarity} respectively; a description of the datasets and experimental set-up is described in Section \ref{sec3}; obtained results are explained in Section \ref{sec4}, and in Section \ref{sec5}, we conclude the paper with main results and future extensions of the work.

\section{Proposed model}\label{sec2}
\subsection{ARIMAX model}
ARIMA model is a classical time series forecasting model characterized by the parameters $p$, $d$ and $q$. Order of AR and MA models are denoted by $p$ and $q$ respectively. $d$ represents the required number of differencing to make the given time series stationary \cite{chatfield2016analysis}. ARIMA( $p$, $d$, $q$) model is represented as follows:
\begin{eqnarray*}
{y}_t & = & \theta_0  + \phi_1 y_{t-1} + \phi_2 y_{t-2} + \cdots +
\phi_py_{t-p} + \varepsilon_{t}   - \theta_1\varepsilon_{t-1} -
\theta_2\varepsilon_{t-2} - \cdots -
          \theta_q\varepsilon_{t-q},
\end{eqnarray*}
where the observed time series is denoted by $y_t$, $\varepsilon_{t}$ is the random error at time $t$, $\phi_i$ are the AR coefficients and $\theta_j$ are the MA coefficients \cite{chatfield2016analysis}.

ARIMAX model simply adds in the exogenous variable $x_t$  at the right hand side as follows

\begin{eqnarray*}
{y}_t & = & \beta x_t + \theta_0  + \phi_1 y_{t-1} + \phi_2 y_{t-2} + \cdots +
\phi_py_{t-p} + \varepsilon_{t}   - \theta_1\varepsilon_{t-1} -
\theta_2\varepsilon_{t-2} - \cdots -
          \theta_q\varepsilon_{t-q},
\end{eqnarray*}

where $\beta$ is the coefficient of the exogenous variable \cite{hyndman2018forecasting}.

\subsection{ARNN model} 
\begin{figure}
    \centering
    \includegraphics[width = 0.85\textwidth]{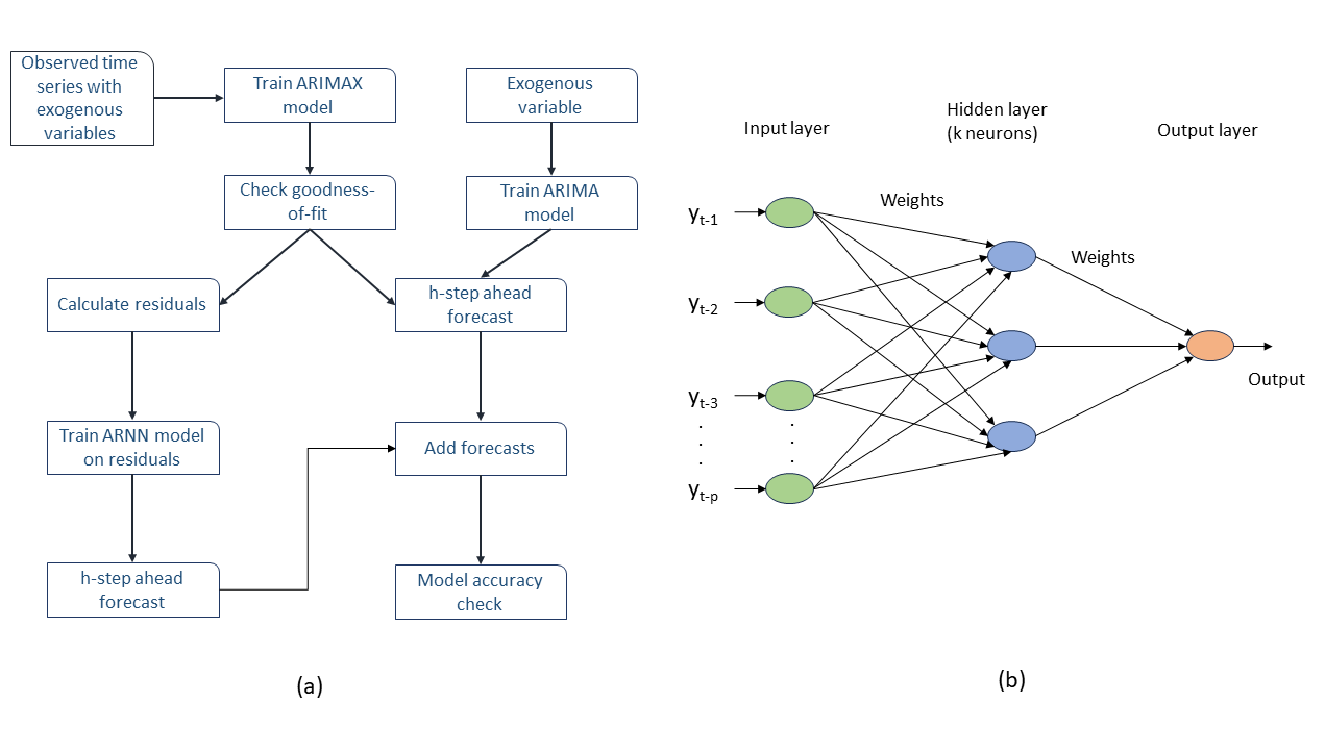}
    %\centering
    \caption{(a) Flow diagram of the proposed model. (b) ARNN model structure.}
    \label{proposed_model_diagram}
\end{figure}

The ARNN model is a feed-forward neural network which takes lagged values of the given time series as input. Feed-forward neural networks 
are simplified mathematical models to represent a living brain. These networks are used to tackle complex modelling and forecasting problems.
The units of a network are neurons that are arranged in multiple layers: usually input, hidden and output layers (as displayed in Fig. \ref{proposed_model_diagram}(b)). A learning algorithm is employed to minimize the sum of squared errors between a given series and output from the neural network model. The associated coefficients are estimated while fitting the model to the given series. The ARNN model is characterized by  two parameters, $p$ and $k$. $p$ is the AR coefficients obtained by calculation of AIC values for different choices of $p$. $k$ is the number of nodes in the hidden layer which is defined by setting $k=[\frac{(p+1)}{2}]$ for non-seasonal data sets. ARNN model is applied iteratively with the logistic activation function within the network while calculating projections. The model uses a moving window approach for making multi-step ahead forecasts in a time series.

\subsection{ARIMAX-ARNN model}
Both ARIMAX and ARNN models have different capabilities to capture relatively disjoint characteristics of the data. A hybrid approach can thus model complex patterns to improve the overall forecasting performance. The underlying assumption of this hybrid approach is that the relationship between two disjoint components is additive. The hybrid model can then be represented as:
$$Z_{t} = Y_{t} + N_{t}$$
where $Y_{t}$ is the simple linear part and $N_{t}$ is the remaining complex part of the model. Let $\hat{Y_{t}}$ be the forecast value of the ARIMAX model at time $t$ and $\epsilon_t$ represent the residual at time $t$ as obtained from the ARIMAX model; then
$$\epsilon_t = Z_{t} - \hat{Y_{t}}$$
The residuals are then modeled by the ARNN model that can be represented as:
$$\epsilon_t = f(\epsilon_{t-1},\epsilon_{t-2},...,\epsilon_{t-n}) + \zeta_t $$
where f is a non-linear function modeled by the ARNN approach and $\zeta_t$ is the random error. The combined forecast becomes:
$$\hat{Z_{t}} = \hat{Y_{t}} + \hat{N_{t}}$$
where $\hat{N_{t}}$ is the forecast value of the ARNN model. The rationale behind fitting a ARNN model on the residuals is that the complex non-linear auto-correlation left in the residuals that the ARIMAX failed to capture may be captured by the ARNN model. The proposed hybrid ARIMAX-ARNN model works in two phases, in the first phase an ARIMAX model analyzes the simple linear part of the model, and in the next stage, an ARNN model is employed to model the residuals of the ARIMAX model. The algorithm can be outlined as follows:

\begin{enumerate}[label={(\alph*)}]
\item {Given a time series of length $n$, input the training dengue incidence data.}
\item {Determine the best ARIMAX model using the training data consisting of dengue incidence and rainfall data.}
\item Train the residual series $(\varepsilon_{t})$ generated by ARIMAX by the ARNN model, as described in Section \ref{sec2}.
\item Obtain the required number (h-step ahead) of forecast of the exogenous variable using ARIMA model.
\item{Using forecasted values of the exogenous variable and the best fitted ARIMAX model, generate the h-step ahead forecasts.}
\item Generate an h-step ahead forecast using a fitted ARNN model of the residuals.
\item Compare the forecasting capabilities of the candidate model by calculating RMSE and MAE in the test period.
\item Final predictions $(\hat Y_t)$ are the obtained by combining
ARIMAX predictions with ARNN predictions $(\hat{\varepsilon_{t}})$ for both the training series as well as the out-of-sample forecasts.
\end{enumerate}

\section{Asymptotic Stationarity of the Proposed Model}\label{Sec:Asmp_Stationarity}
This ARIMAX model is, in general, non-stationary. The stationarity of the model can be achieved by the simultaneous stationarity of both the time series and the exogenous variable. Here, we have checked the stationarity of both the weekly rainfall data of Iquitos and San Juan by using the ADF test and found that both are stationary (See table \ref{tab:my_label}).

However, for the analytical purposes, if exogenous or times series or both data are found to be non-stationary, then one way to to achieve the stationarity is differencing $(y_t - \beta x_t)$ up to $d$ times with the appropriate choice of $\beta$. ANNs exhibit asymptotic stationarity, requiring stationarity during the neural network training process. However, their performance falters when applied to nonstationary processes, leading to suboptimal out-of-sample predictions \cite{leoni2009long}. In formulating the hybrid ARIMAX-ARNN model, we will employ an approach analogous to the asymptotic stationarity method proposed by Chakraborty et al. \cite{chakraborty2020unemployment}. Following their method, first we consider a nonlinear ARNN model generated by additive noise of the ARIMAX model. An ARNN process of order $p$ and has $k$ hidden nodes in its one hidden layer i.e. $ARNN(p,k)$ model is of the form  $\epsilon_t=f(\epsilon_{t-1},\epsilon_{t-2},\cdots,\epsilon_{t-p},\psi)+\zeta_t~~\cdots~(1)$

Here, $\epsilon_t$ denotes a time series generated by a nonlinear autoregressive process, $\zeta_t$ is an i.i.d. noise process and $f (.,\psi)$ is a feedforward neural network with
weight parameter vector $\psi$.

Next we consider $f(z)=c_0+\sum_{i=1}^{n}w_i \sigma (a_i+a_i' z)~~\cdots~~(2)$

where $f$ is a neural network, $z$ denotes a p-dimensional input feature to the ARNN model (error residuals obtained from ARIMAX model), $a_i, w_i, c_0$ are scalar weights, and $\sigma$ is a bounded nonlinear sigmoid function. 

We also follow the same idea for the unbounded noise terms, i.e. $z_{t-1}=(\epsilon_{t-1},\cdots,\epsilon_{t-p})'$, $F(z_{t-1})=(f(z_{t-1}),\epsilon_{t-1},\cdots,\epsilon_{t-p})',~e_t=(\zeta_t,0,\cdots,0)'$ and the first-order vector model is as follows.\\
$z_t=F(z_{t-1})+e_t~~\cdots~~(3)$

\begin{theorem}
    Let $E\mid \zeta_t \mid < \infty$ and the PDF of $\zeta_t$ is positive everywhere in $R$, and $\{ \epsilon_t \}$ and $\{z_t \}$ are defined as in $(1)$ and $(3)$, respectively. Then if $f$ is a nonlinear neural network as defined in $(2)$, then $\{z_t\}$ is geometrically ergodic and $\{ \epsilon_t \}$ is asymptotically stationary.
\end{theorem}
\begin{proof}
    The proof is similar and straight forward to Theorem 2 of Chakraborty et. al \cite{chakraborty2020unemployment}. 
\end{proof}

\begin{table}[]
    \centering
    \begin{tabular}{c|c|c}
    \hline
    Data                          & ADF Value   &  p-value \\
    \hline
     San Juan weekly rainfall data &  -8.74   &  $<0.05$ \\
     Iquitos weekly rainfall data &  -6.46  & $<0.05$\\
     \hline
    \end{tabular}
    \caption{Results of Augmented Dickey-Fuller (ADF) test}
    \label{tab:my_label}
\end{table}

\section{Numerical experiments}\label{sec3}
\subsection{Data description}
\begin{figure}
    \centering
    \includegraphics[width=0.48\textwidth]{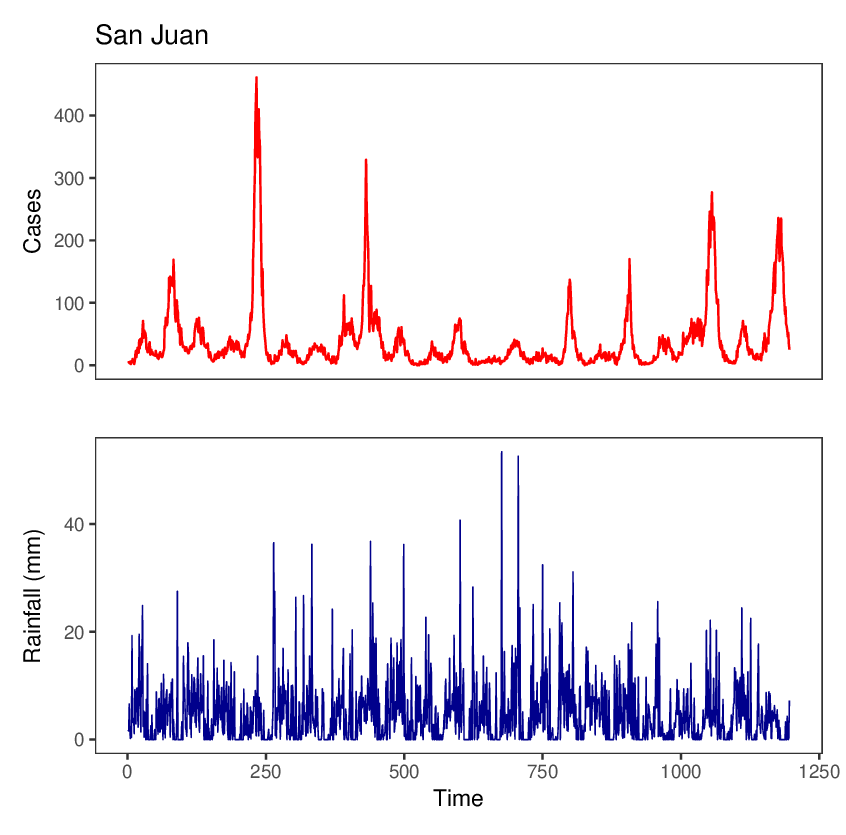}
    \includegraphics[width=0.48\textwidth]{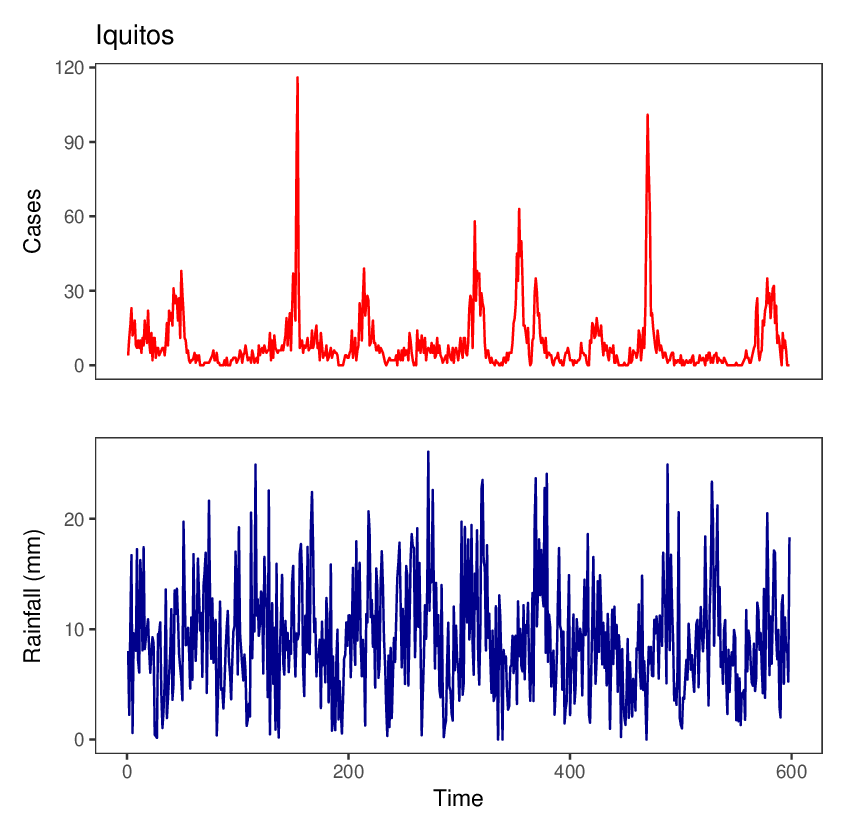}
    \caption{Dengue cases (red) and rainfall (blue) data of San Juan and Iquitos (incidences per 10000 population).}
    \label{fig:data_rainfall}
\end{figure}

Two weekly time series datasets (publicly available at \url{https://dengueforecasting.noaa.gov/}) with an exogenous variable is utilized to test the forecasting capability of a novel hybrid model. The period of the dataset for San Juan and Iquitos are 1990/1991 -- 2008/2009 and 2000/2001 -- 2008/2009 respectively. Several studies have considered many traditional as well as modern machine learning time series models to predict the dengue incidence in San Juan and Iquitos; for instance, see \cite{buczak2018ensemble, johnson2018phenomenological, yamana2016superensemble, chakraborty2019forecasting, deb2022ensemble, panja2023ensemble} and references therein. To check the forecasting capability of the proposed model, we choose rainfall as an exogenous variable \cite{panja2023ensemble}. The average rainfall in San Juan and Iquitos are 4.78 mm (0 -- 53.39 mm) and 9.09 mm (0 -- 26.07 mm) respectively. The dengue incidence and rainfall in these two locations are depicted in Fig. \ref{fig:data_rainfall}. 

\subsection{Performance Metrics} \label{Perf_Met}
Mean Absolute Error (MAE) and Root Mean Square Error (RMSE) are used to compare the performance of the proposed model and the benchmark models. These metrics are defined as follows:
\begin{align}\label{accuracy_metrics}
    MAE &= \frac{\displaystyle 1}{\displaystyle h} \sum_{i=1}^h |y_i - \hat{y}_i|;\\\nonumber
    RMSE &= \sqrt{\frac{1}{h} \sum_{i=1}^h (y_i - \hat{y}_i)^2} ; 
\end{align}

where $y_i$ and $\hat{y}_i$ are the observed cases and their corresponding prediction, and $h$ is the forecast horizon. It is general convention that the model with the least accuracy metric values is the ``best" among competing models \cite{panja2023ensemble}.

\subsection{Data splitting}
\begin{figure}
    \centering
    \includegraphics[width = 1.0\textwidth]{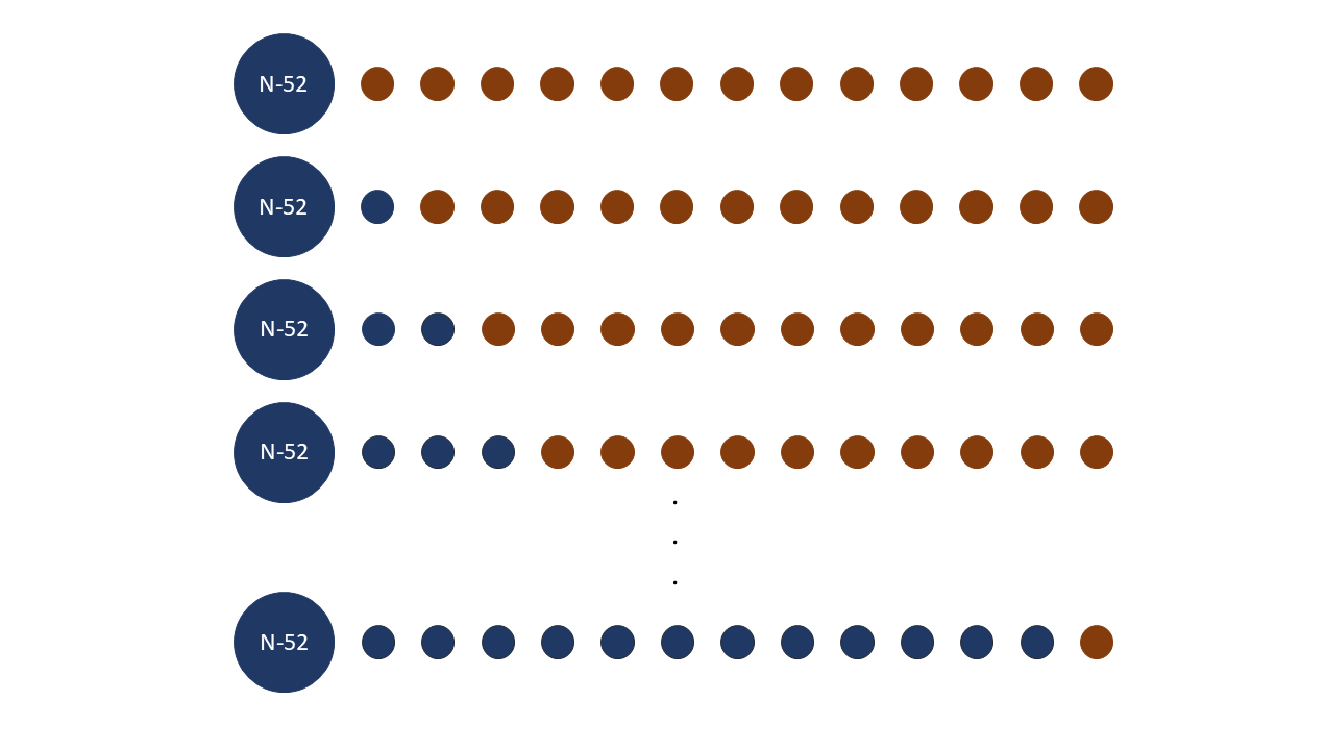}
    %\centering
    \caption{Splitting of the datasets for experiments. Each small circle represents 4 weeks of data. Blue circles denote training data and brown circles denote test data.}
    \label{fig:data_split}
\end{figure}

To evaluate the forecasting capability of the proposed model we use varying training sets and test sets. We gradually increase the size of the training set to check the robustness of the models. We argue that a robust model will show increasing accuracy of forecasts (i.e., decreased values of MAE and RMSE) as we increase the training period. When there are N weeks of data, we take (N-52) as the training set and the rest of the dataset as the test set. Subsequently, the training set is increased by a four-week period in each simulation. A schematic diagram of the splitting is depicted in Fig. \ref{fig:data_split}.

\subsection{Benchmark models and their implementation} \label{benchmarks}
We compute the forecasting performances of benchmark models that are used in multivariate forecasting tasks on the two dengue datasets with rainfall as an exogenous variable. The objective is to check the forecasting capability of the proposed model compared to the benchmark models based on accuracy metrics. All the competing models will be trained on historical dengue cases and rainfall data. We employ the following state-of-the-art models as benchmarks - auto-regressive integrated moving average with exogenous variables (ARIMAX) \cite{hyndman2018forecasting}, exponential smoothing with exogenous variables (ETSX) \cite{hyndman2018forecasting}, auto-regressive neural network with exogenous variables (ARNNX) \cite{hyndman2018forecasting};  Block recurrent neural network with exogenous variables (BlockRNNX) \cite{herzen2022darts}, Neural basis expansion analysis with exogenous variables (NBeatsX) \cite{NBEATS},  Transformers with exogenous variables (TransformersX) \cite{TRANSFORMERS} and Temporal convolutional networks with exogenous variables (TCNX) \cite{TCN}. 

The proposed hybrid ARIMAX-ARNN model is implemented in R software. The dengue cases and rainfall datasets are divided into training and test sets with lengths defined in Fig. \ref{fig:data_split}. Variable test sets are considered for checking the robustness of the results. Appropriate ARIMAX model is estimated for each training data using the in-built R function "auto.arima" in "forecast" package \cite{hyndman2020package}. The residuals are then calculated for each training sets. On each residual series, ARNN model is built using "nnetar" function in "forecast" package \cite{hyndman2020package}. See Section \ref{sec2} for more details on the implementation of ARNN model. From both ARIMAX and ARNN model, h-step-ahead forecasts are generated (h is the length of the test set). For the h-step-ahead forecasts, h values of the rainfall is also required. To do this, we fit a separate ARIMA model on the rainfall data using "auto.arima" function and generate h-step-ahead forecast of rainfall to be used as regressors in the test period. The codes are available in the GITHUB link \url{https://github.com/ShashankGupta01/Exogenous}. Benchmark forecasting candidates such as ARIMAX, ETSX, ARNNX are also implemented using "forecast" package. The deep learning models are implemented in Python using "darts" package \cite{herzen2022darts}.

\section{Results}\label{sec4}
\begin{figure}
    \centering
    \includegraphics[width = 1.0\textwidth]{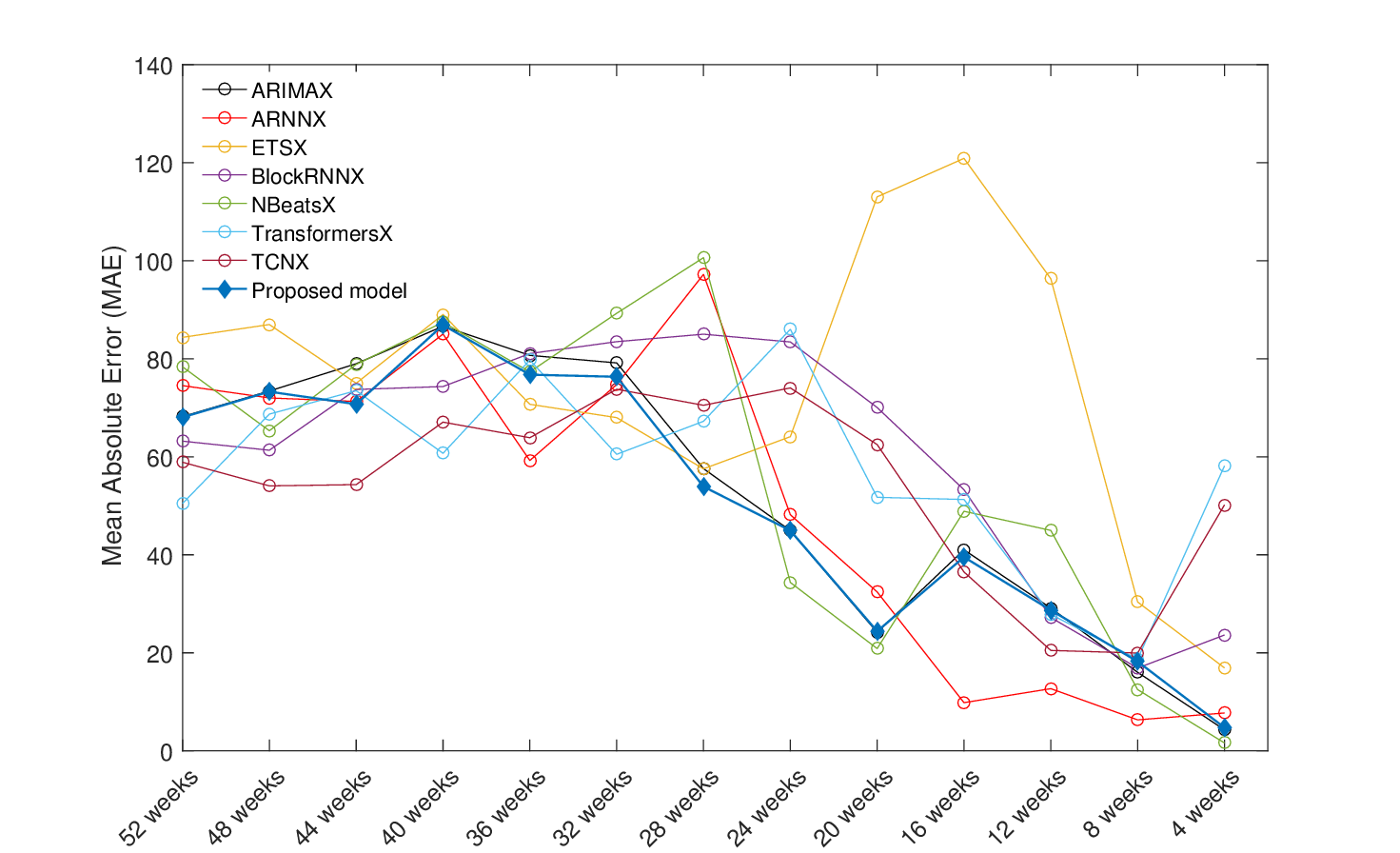}
    \includegraphics[width = 1.0\textwidth]{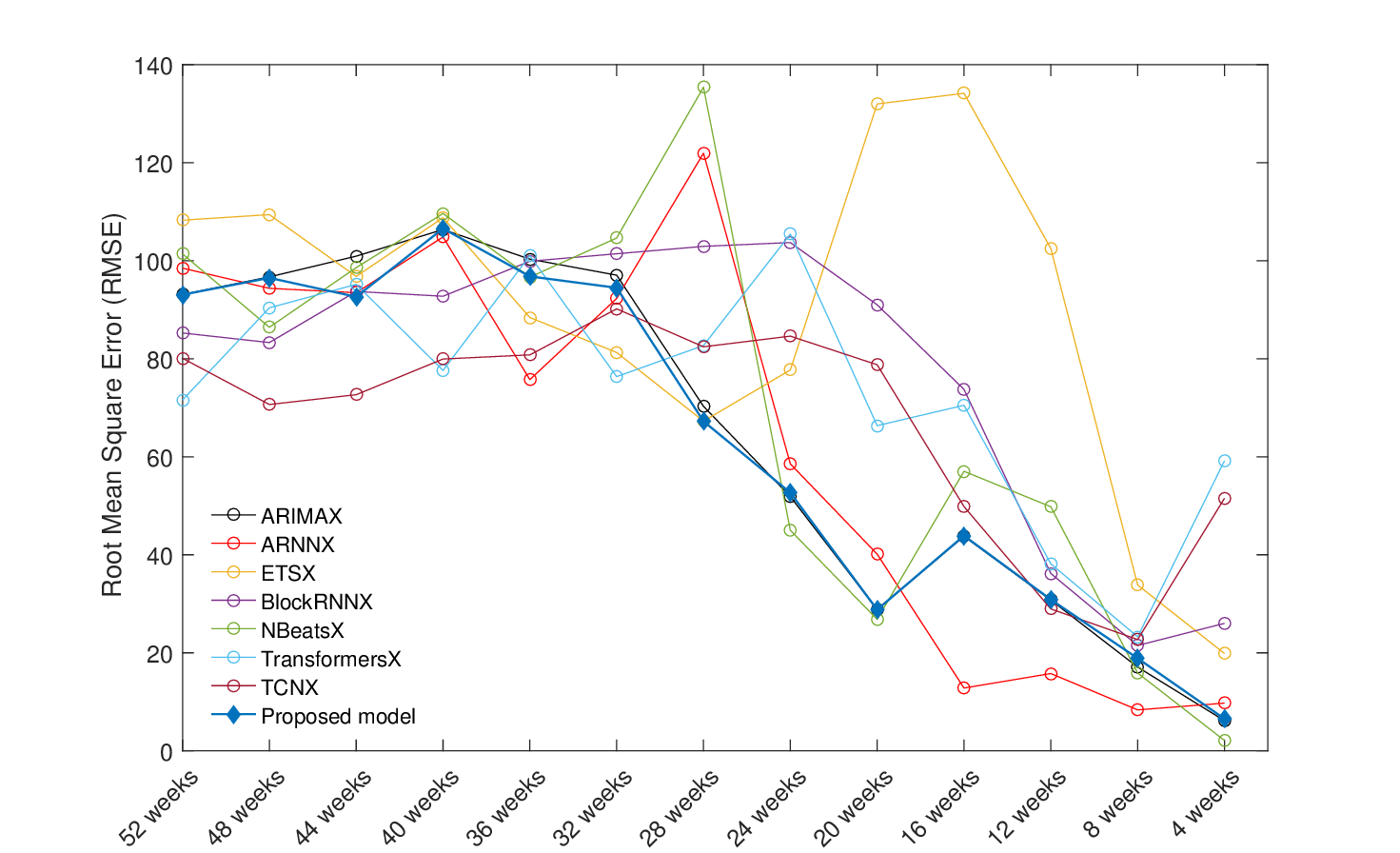}
    %\centering
    \caption{Fitting results for San Juan over the gradually decreasing length of test data. The upper panel shows changes in MAE and the lower panel depicts changes in RMSE  for the 8 competing models.}
    \label{fig:sanjuan_results}
\end{figure}

The accuracy metrics for the San Juan data are depicted in Fig. \ref{fig:sanjuan_results}. For 52 weeks test data of San Juan, TransformerX outperforms all other models in both accuracy metrics MAE and RMSE. However, TransformerX does not show consistency when adding more data to the training set. Among the statistical models, ARNNX and ARIMAX perform significantly better than the ETSX model for San Juan. It can be observed that the proposed model improves ARIMAX forecasts in the 28 weeks, 32 weeks and 36 weeks forecast horizons. This indicates that the proposed model shows comparable forecasting accuracy and consistency for San Juan data.
 
\begin{figure}
    \centering
    \includegraphics[width = 1.0\textwidth]{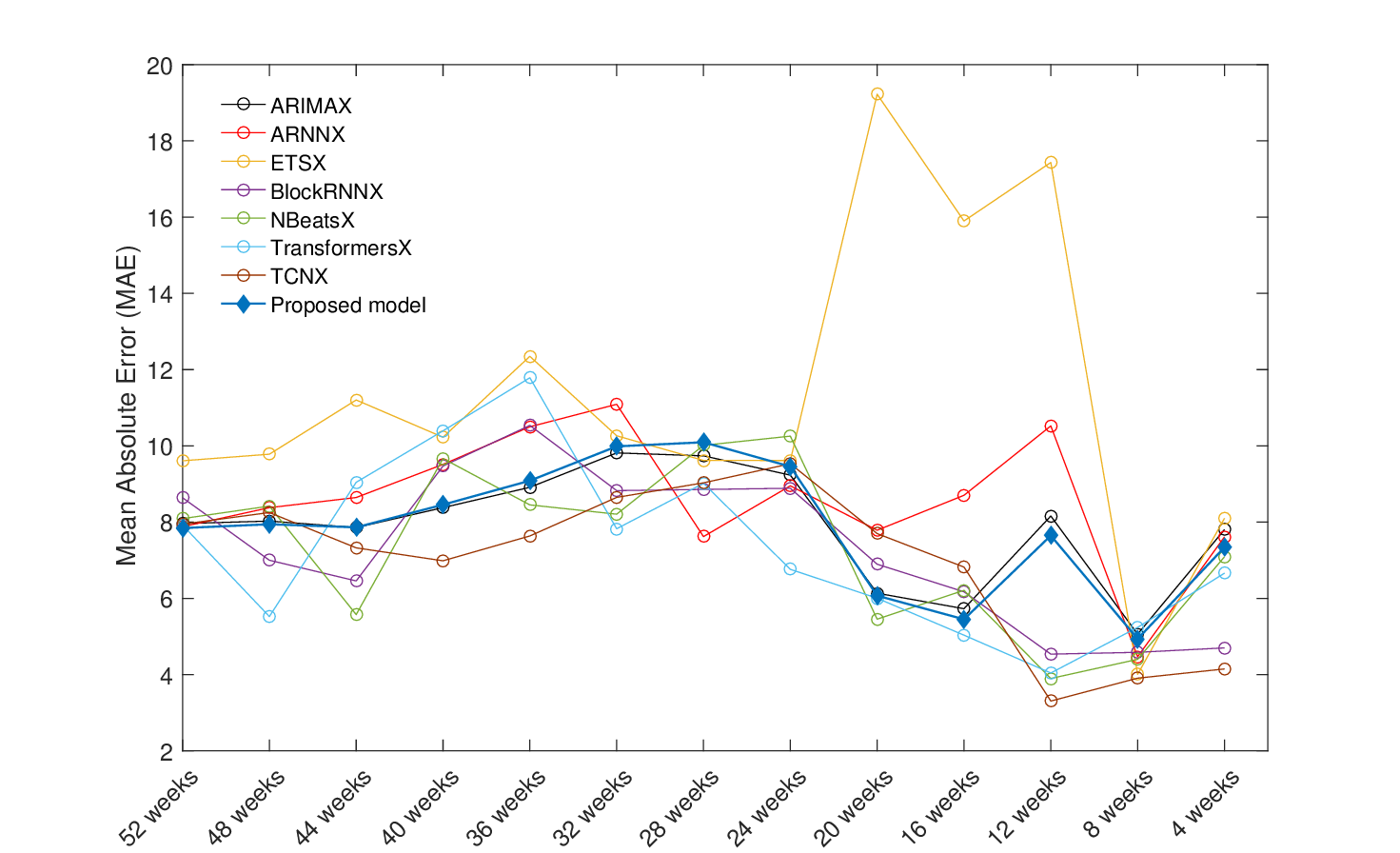}
    \includegraphics[width = 1.0\textwidth]{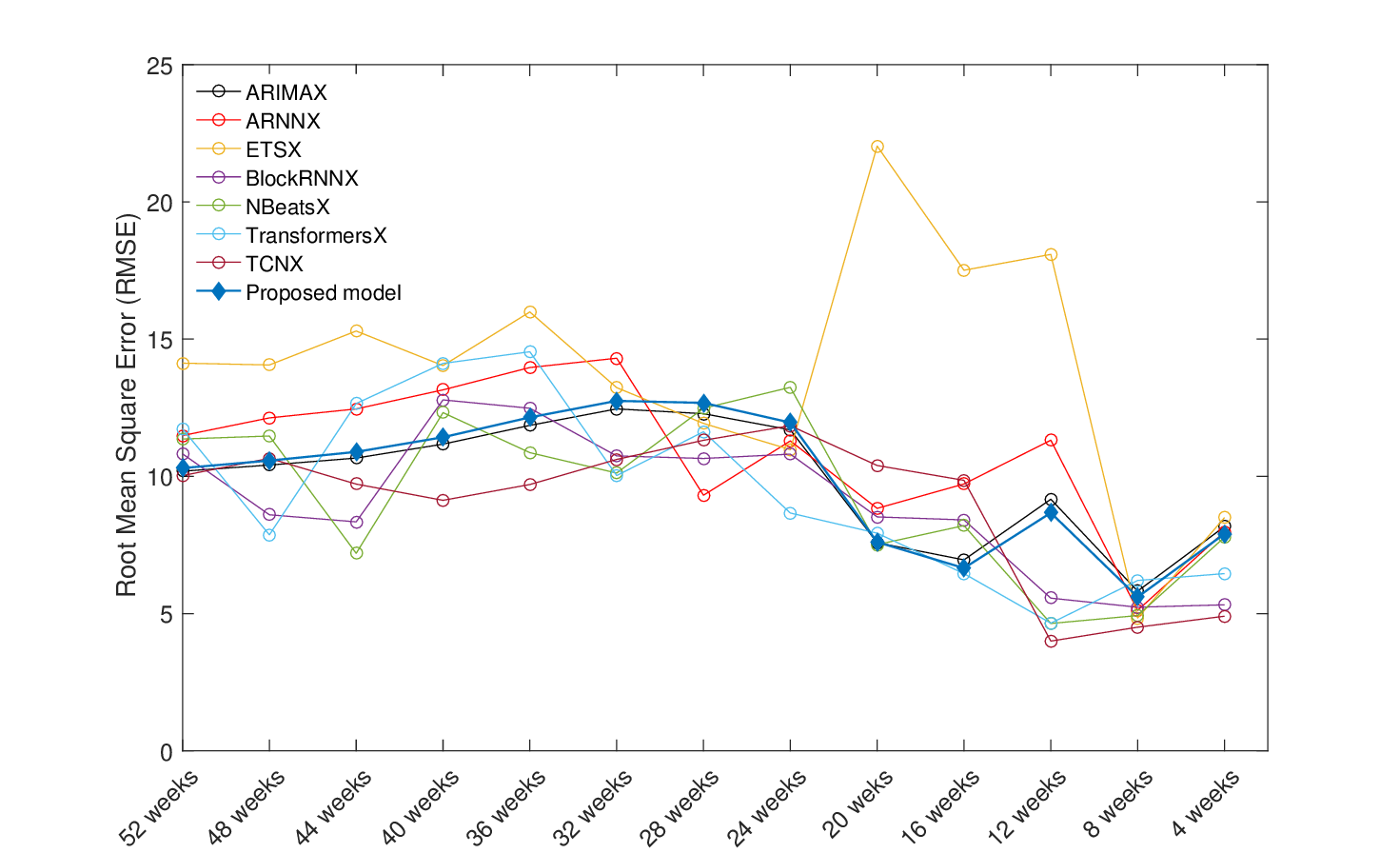}
    %\centering
    \caption{Fitting results for Iquitos over the gradually decreasing length of test data. The upper panel shows changes in MAE and the lower panel depicts changes in RMSE for the 8 competing models.}
    \label{fig:Iquitos_results}
\end{figure}

The accuracy metrics for the Iquitos data are depicted in Fig. \ref{fig:Iquitos_results}. For Iquitos data, ETSX performs poorly compared to the ARIMAX and ARNNX models. Proposed model improves ARIMAX forecasts in a shorter forecasting horizon. Among deep learning models, TCNX and BlockRNNX perform better than TCNX and TransformerX in most forecast horizons. Before 20 weeks forecasting horizon, ARIMAX performs better than the proposed model. However, the proposed improves the ARIMAX forecasts on or after 20 weeks horizon in Iquitos. This indicates that the proposed model performs better than ARIMAX in short-term forecasting horizons in the Iquitos dataset. 

From the results of the benchmark forecasting models and the proposed model on the two dengue datasets, a few interesting observations are revealed. Classical ARIMAX model and neural network based ARNNX model are showing very competitive results despite being simpler than other competing models. Deep learning based models show higher accuracy than other models in sporadic situations. However, these models are mostly inconsistent. The reason behind this may be the length of the two dengue datasets are limited as the deep learning models need much longer historical data. However, the proposed model improves the forecasts of ARIMAX in some forecast horizons making it a potential alternative to the ARIMAX model. Our main objective is to propose a multivariate extension of hybrid ARIMA-ARNN model. The results also show that the proposed hybrid ARIMAX-ARNN model will be an important tool to tackle dengue forecasting problem with exogenous variables.

\section{Discussion and conclusion}\label{sec5}
Dengue forecasting is a challenging problem which requires the integration of various data sources. In this paper, we propose the multivariate extension of the hybrid ARIMA-ARNN model with dengue incidence data as input and rainfall as an exogenous variable. We utilized publicly available data from San Juan and Iquitos to test and compare the forecasting capability of the proposed model with some state-of-the-art benchmark models. It has been observed that the proposed model shows competitive forecasting accuracy in variable forecast horizons. Moreover, the proposed model is shown to achieve asymptotic stationarity using a previously established theorem. 

However, we build the model by understanding the components and then explained the hybridization (see Fig. \ref{proposed_model_diagram}). Several state-of-the-art benchmark models are employed to forecast the same dengue datasets with gradually increasing training dataset. We adopt the increasing training data and decreasing test sets to compare the robustness of the competing models. Two accuracy metrics, namely, MAE and RMSE are used to compare the models' outputs. Accuracy metrics for the gradually increasing training sets indicate that ARIMAX and ARNNX models are most consistent (see Fig. \ref{fig:sanjuan_results} and Fig. \ref{fig:Iquitos_results}). Furthermore, the proposed model improves ARIMAX forecasts in some forecasting horizons and lie very close to the ARIMAX results otherwise. Deep learning models are inconsistent as long as all the forecast horizons are considered due to the short length data sets. 

Further research may include more exogenous variables, such as temperature and other socio-economic variables, in the proposed model. The proposed model can be readily used in other disease forecasting tasks with exogenous variables. Moreover, the proposed model can also be used in other forecasting problems such as stock markets, weather, ecological variables, oil price prediction and wherever there is a forecasting task with exogenous variables. 

In summary, the proposed hybrid ARIMAX-ARNN model is a plausible extension of the hybrid ARIMA-ARNN model that can be used as a potential competitor for dengue forecasting tasks with exogenous variables. Numerical experiments reveal the consistency of the proposed model and its usability as an early warning system for dengue outbreaks.

%\section*{Acknowledgements}

\bibliographystyle{plain}
\biboptions{square}
\bibliography{bibliography}

\end{document}